\documentclass[graybox]{svmult}

\usepackage[latin1]{inputenc}
\usepackage[T1]{fontenc}
\usepackage{amsmath}
\usepackage{amssymb}
\usepackage[footnotesize]{caption}
\usepackage{ifpdf}

\ifpdf
\usepackage[pdftex]{graphicx}
\usepackage[pdftex]{hyperref}
\else
\usepackage[dvips]{graphicx}
\newcommand{\href}[2]{#2}
\fi

\newcommand{\al}{\alpha}

\newcommand{\gam}{\gamma}
\newcommand{\Gam}{\Gamma}
\newcommand{\del}{\delta}
\newcommand{\Del}{\Delta}
\newcommand{\eps}{\varepsilon}

\newcommand{\tht}{\theta}
\newcommand{\lam}{\lambda}

\newcommand{\sig}{\sigma}
\newcommand{\vphi}{\varphi}

\newcommand{\Om}{\Omega}

\newcommand{\ket}[1]{\left| #1 \right \rangle}
\newcommand{\ens}[1]{\left \langle #1 \right \rangle}

\newcommand{\hatn}{\mathbf{\hat{n}}}

\newcommand{\ergo}{\Rightarrow}

\renewcommand{\L}{\mathcal{L}}

\newcommand{\GeV}{\textrm{GeV}}
\newcommand{\cm}{\textrm{cm}}
\renewcommand{\sec}{\textrm{sec}}
\newcommand{\TeV}{\textrm{TeV}}
\newcommand{\keV}{\textrm{keV}}
\newcommand{\MeV}{\textrm{MeV}}

\def\fun#1#2{\lower3.6pt\vbox{\baselineskip0pt\lineskip.9pt
 \ialign{$\mathsurround=0pt#1\hfil##\hfil$\crcr#2\crcr\sim\crcr}}}
\def\lesssim{\mathrel{\mathpalette\fun <}}
\def\gtrsim{\mathrel{\mathpalette\fun >}}

\usepackage{mathptmx}       % selects Times Roman as basic font
\usepackage{helvet}         % selects Helvetica as sans-serif font
\usepackage{courier}        % selects Courier as typewriter font
\usepackage{type1cm}        % activate if the above 3 fonts are
\usepackage{makeidx}         % allows index generation
\usepackage{graphicx}        % standard LaTeX graphics tool
\usepackage{multicol}        % used for the two-column index
\usepackage[bottom]{footmisc}% places footnotes at page bottom

\makeindex             % used for the subject index
\begin{document}

\title*{Dark Matter Astrophysics}
% Use \titlerunning{Short Title} for an abbreviated version of
% your contribution title if the original one is too long
\author{Guido D'Amico, Marc Kamionkowski and Kris Sigurdson}
% Use \authorrunning{Short Title} for an abbreviated version of
% your contribution title if the original one is too long
\institute{Guido D'Amico \at SISSA, via Beirut 2-4, 34014
Trieste, Italy; \email{damico@sissa.it} 
\and Marc Kamionkowski \at
California Institute of Technology, Mail Code 350-17, Pasadena,
CA 91125, USA; \\ \email{kamion@caltech.edu} 
\and Kris Sigurdson \at
Department of Physics and Astronomy, University of British
Columbia, Vancouver, BC V6T 1Z1, Canada; \email{krs@phas.ubc.ca}}
%
% Use the package "url.sty" to avoid
% problems with special characters
% used in your e-mail or web address
%
\maketitle

\abstract*{
These lectures are intended to provide a brief pedagogical
review of dark matter for the newcomer to the subject.  We begin
with a discussion of the astrophysical
evidence for dark matter.  The standard weakly-interacting
massive particle (WIMP) scenario---the motivation, particle
models, and detection techniques---is then reviewed.  We
provide a brief sampling of some recent variations to the
standard WIMP scenario as well as some alternatives (axions and
sterile neutrinos).  Exercises are provided for the reader.}

\abstract{These lectures are intended to provide a brief pedagogical
review of dark matter for the newcomer to the subject.  We begin
with a discussion of the astrophysical
evidence for dark matter.  The standard weakly-interacting
massive particle (WIMP) scenario---the motivation, particle
models, and detection techniques---is then reviewed.  We
provide a brief sampling of some recent variations to the
standard WIMP scenario as well as some alternatives (axions and
sterile neutrinos).  Exercises are provided for the reader.\footnote{Based on lectures
given by MK at the Villa Olmo School on ``The Dark Side of the
Universe,'' 14--18 May 2007 and by KS at the XIX Heidelberg
Physics Graduate Days, 8-12 October 2007.}}

\section{Introduction}

Dark matter is an essential ingredient in a good fraction of the
literature on extragalactic astronomy and cosmology.
Since dark matter cannot be made of any of the usual
standard-model particles (as we will discuss below), dark matter
is also a central focus of elementary-particle physics.  The purpose of
this review is to provide a pedagogical introduction to
the principle astrophysical evidence for dark matter and to
some of the particle candidates.

Rather than present a comprehensive survey of the vast and
increasingly precise measurements of the amount and distribution
of dark matter, we will present very simple (``squiggly-line'')
arguments for the existence of dark matter in clusters and
galaxies, as well as the arguments for why it is nonbaryonic.
The motivation will be to provide insight into the evidence and
arguments, rather than to summarize results from the latest
state-of-the-art applications of the techniques.  

Likewise, construction of particle-physics models for dark
matter has become a huge industry, accelerated quite recently, in
particular, with anomalous cosmic-ray and diffuse-background
results \cite{haze,pamela}.  Again, we will
not attempt to survey these recent developments and focus
instead primarily on the basic arguments for particle dark
matter.  In particular, there has developed in the theoretical
literature over the past twenty years a ``standard''
weakly-interacting massive particle (WIMP) scenario, in which
the dark-matter particle is a particle that arises in extensions
(e.g., supersymmetry \cite{Jungman:1995df} or universal extra
dimensions \cite{UEDs}) of the
standard model that are thought by many particle theorists to
provide the best prospects for new-physics discoveries at the
Large Hadron Collider (LHC).  We therefore describe this basic
scenario.  More detailed reviews of weakly-interacting massive
particles, the main subject of this article, can be found in
Refs.~\cite{Jungman:1995df,Bergstrom:2000pn,Bertone:2004pz}.

After describing the standard WIMP scenario, we provide a brief
sampling of some ideas for ``non-minimal'' WIMPs, scenarios in
which the WIMP is imbued with some additional properties, beyond
simply those required to account for dark matter.  We also
briefly discuss some other attractive ideas (axions and sterile
neutrinos) for WIMPs.  Exercises are provided throughout.

\section{Astrophysical evidence}

It has been well established since the 1930s that there is much
matter in the Universe that is not seen.  It has also been long
realized, and particularly since the early 1970s, that much of
this matter must be nonbaryonic.  The evidence for a significant
quantity of dark matter accrued from galactic dynamics, the
dynamics of galaxy clusters, and applications of the cosmic
virial theorem.  The evidence that much of this matter is
nonbaryonic came from the discrepancy between the total matter
density $\Omega_m\simeq 0.2-0.3$ (in units of the critical density $\rho_c=3
H_0^2/8\pi G$, where $H_0$ is the Hubble parameter), obtained from
such measurements, and the baryon density $\Omega_b\simeq 0.05$
required for the concordance between the observed light-element (H,
D, $^3$He, $^4$He, $^7$Li) abundances with those predicted by
big-bang nucleosynthesis \cite{Iocco:2008va}, the theory for the
assembly of light elements in the first minutes after the big bang.

Rather than review the historical record, we discuss the most
compelling arguments for nonbaryonic dark matter today as well
as some observations most relevant to astrophysical
phenomenology of dark matter today.

\subsection{Galactic rotation curves}

The flatness of galactic rotation curves has
provided evidence for dark matter since the 1970's.
These measurements are
particularly important now not only for establishing the
existence of dark matter, but particularly for fixing the local
dark-matter density, relevant for direct detection of dark matter.  We
live in a typical spiral galaxy, the Milky Way, at a distance
$\sim 8.5 \textrm{ kpc}$ from its center.  The visible stars and
gas in the Milky Way extend out to a distance of about 10 kpc.
From the rotation curve, the rotational velocity
$v_c(r)$ of stars and gas as a function of Galactocentric radius
$r$, we can infer the mass $M_{<}(r)$ of the Galaxy enclosed within
a radius $r$.  If the visible stars and gas provided all the
mass in the Galaxy, one would expect that the
rotation curve should decline at radii larger than the 10~kpc
extent of the stellar disk according to the
Keplerian relation  $v_c^2 = G M_{obs}/r$.  Instead, one
observes that $v_c(r)$ remains constant (a flat rotation curve)
out to much larger radii, indicating that $M_{<}(r) \propto r$ 
for $r \gg 10~{\rm kpc}$ and thus that the Galaxy must contain
far more matter than contributed by the stars and the gas.

\begin{figure}[b]
  \centering
  \includegraphics[scale=0.5]{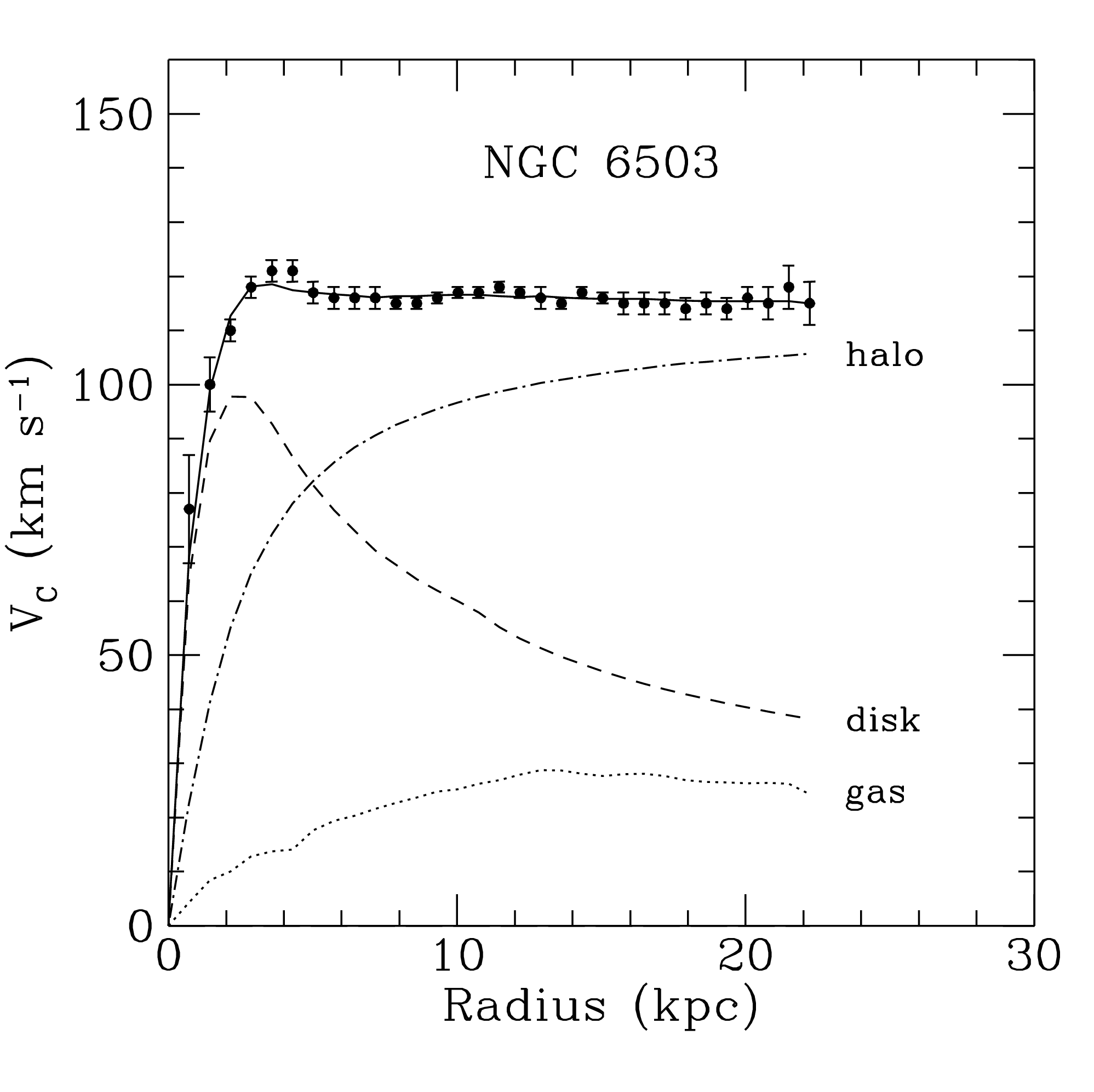}
  \caption{Measured rotation curve of NGC6503 with best fit and
  contributions from halo, disk and gas. From
  Ref.~\protect\cite{Begeman91}}
	\label{fig:rotationcurve}
\end{figure}

Assuming a spherically symmetric distribution of matter, the
mass inside a radius $r$ is given by
\begin{equation}
      M_{<}(r) = 4 \pi \int_0^r \rho(r') r'^2 dr'.
\end{equation}
An estimate for the distribution of dark matter in the Galaxy
can be obtained from the behavior of the rotation curve in the
inner and outer galaxy.
For example, the density distribution for the cored isothermal sphere,
given by,
\begin{equation}
\label{isothermal}
     \rho(r) = \rho_0 \frac{R^2 + a^2}{r^2 + a^2} \, ,
\end{equation}
where $R \sim 8.5~{\rm kpc}$ is our distance from the Galactic
center and $\rho_0$ is the local dark-matter density, provides a
qualitatively consistent description of the data.  For large
$r$, $\rho \sim r^{-2} \, \ergo \, M(r) \propto r \, \ergo \, v
\sim {\rm const}$, while for small $r$, $\rho \sim
\textrm{const} \, \ergo \, M(r) \propto r^3 \, \ergo \, v
\propto r$.  Eq.~(\ref{isothermal}) describes a 2-parameter
family of density profiles and by fitting the observed data one
finds a scale radius $a \sim 3-5 \textrm{ kpc}$ and local matter
density $\rho_0 \sim 0.4 \, \GeV \, \cm^{-3}$; the uncertainties
arise from standard error in the rotation-curve measurements and
from uncertainties in the contribution of the stellar disk to
the local rotation curve.  Because the dark
matter is moving in the same potential well, the velocity
dispersion of the dark matter can be estimated to be
$\left\langle v_{{\rm dm}}^2 \right\rangle^{1/2} \sim 300~{\rm
km}/{\rm sec}$.  The simplest assumption is that the dark matter
has a Maxwell-Boltzmann distribution with $f(\vec{v}) \sim e^{-
v^2/2 \bar v^2}$, where $\bar v \sim 220\,  {\rm km}/{\rm sec}$. 

\medskip
\noindent {\sl Exercise 1. Explain/estimate how $\rho_0$ would
be affected if
\begin{itemize}
 \item (a) the halo were flattened, keeping the rotation curve unaltered;
 \item (b) the profile were of the Navarro-Frenk-White (NFW) type:
 $\rho(r) \propto \rho_c/[r (r + r_c)^2]$, keeping the local
 rotation speed the same;
 \item (c) the stellar contribution to the rotation curve was
 either increased or decreased.
\end{itemize}
}
\medskip

\subsection{Galaxy Clusters}

Galaxy clusters are the largest gravitationally bound objects in
the Universe.  They were first observed as concentrations of
thousands of individual galaxies, and early application of the
virial theorem $v^2\sim GM/R$ (relating the observed velocity
dispersion $v^2$ to the observed radius $R$ of the cluster) 
suggested that there is more matter in clusters than the stellar
component can provide \cite{Zwicky:1933}.   It was later
observed that these galaxies are embedded in hot x-ray--emitting
gas, and we now know that clusters are the brightest
objects in the x-ray sky.  The x rays are produced by hot gas
excited to virial temperatures $T \sim \keV$ of the
gravitational potential well of the dark matter, galaxies, and
gas. A virial temperature $T \sim \keV$ corresponds to a
typical velocity for the galaxies of $v \sim 10^3\,
\textrm{km/s}$.

Observations of clusters come from optical and x-ray telescopes
and more recently via the Sunyaev-Zeldovich effect
\cite{Sunyaev:1980vz}.
Several independent lines of evidence from clusters
indicate that the total mass required to explain observations
is much larger than can be inferred by the observed baryonic
content of galaxies and gas.

\subsubsection{Lensing}

\begin{figure}[b]
  \centering
  \includegraphics[scale=0.5]{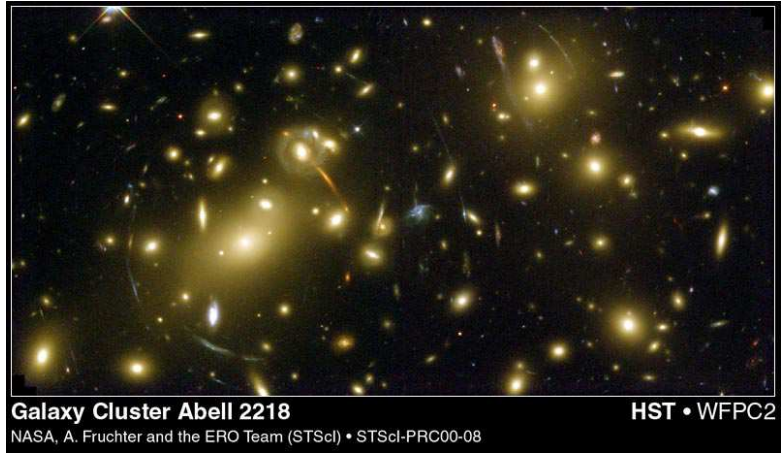}
  \caption{Image of the galaxy cluster Abel 2218.  Credits:
  NASA, Andrew Fruchter and the ERO team.}
	\label{fig:abell2218}
\end{figure}

Galaxy clusters exhibit the phenomenon of  gravitational lensing
\cite{Einstein:1936,Zwicky:1937}.  Because the
gravitational field of
the cluster curves the space around it, light rays emitted from
objects behind the cluster travel along curved rather than
straight paths on their way to our telescopes \cite{Blandford:1991xc}.
If the lensing
is strong enough, there are multiple paths from the same object,
past the cluster, that arrive at our location in the Universe;
this results in multiple images of the same object (e.g., a
background galaxy or active galactic nucleus).  Furthermore,
because the light from
different sides of the same galaxy travels along slightly
different paths, the images of strongly lensed sources are
distorted into arcs.  For instance, HST observations of Abell
2218 show arcs and multiple images as shown in Fig.~\ref{fig:abell2218}.
If the lensing is weak, the images may become slightly
elongated, even if they are not multiply imaged.

For a lensing cluster with total mass $M$ and  impact
parameter $d$ the deflection angle is of order
\begin{equation}
\label{defangle}
     \al \sim \left( \frac{G M}{d c^2} \right)^{1/2}.
\end{equation}
Thus, from measurements of the deflection angle and impact
parameter (which can be inferred by knowing the redshift to the
lensing cluster and source), one can infer that the total mass
$M$ of a cluster is much larger than the observed baryonic
mass $M_b$.

\medskip
\noindent {\sl Exercise 2. Suppose a massive particle with
velocity $v$ is incident, with impact parameter $b$, on a fixed
deflector of mass $M$.  Calculate the deflection angle (using
classical physics) due to scattering of this particle via
gravitational interaction with the deflector.  Show that you
recover $\al = \left( G M/d c^2\right)^{1/2}$ in the limit
$v\rightarrow c$, the velocity at which light rays propagate.
Actually, the correct general-relativistic calculation recovers
this expression, but with an extra factor of 2.}

\medskip
\noindent {\sl Exercise 3. Estimate the deflection angle $\al$
for lensing by a cluster of $M \sim 10^{15} M_{\odot}$ and for
an impact parameter of 1 Mpc.}
\medskip

\subsubsection{Hydrostatic equilibrium}

In a relaxed cluster, the temperature profile
$T(r)$ of gas, as a function of radius $r$, can be inferred
using the strength of the emission lines,
and the electron number density $n_e(r)$ can be inferred
using the the x-ray luminosity $L(r)$.    Combined, these
observations give an estimate of the radial pressure profile
$p(r) \propto n_e(r) k_B T(r)$.   In steady state, a gravitating
gas will satisfy the equation of hydrostatic equilibrium,
\begin{align}
\label{euler}
    \frac{d p}{d r} = - G \, \frac{M_{<}{(r)} \, \rho_{{\rm gas}}(r)}{r^2}\,.
\end{align}
 Here, $M_<(r)$
is the total (dark matter and baryonic gas) mass enclosed by a
radius $r$ and $\rho_{\rm gas}(r)$ is the density at radius
$r$. Eq.~(\ref{euler}) can be used to determine the total mass
$M$ of the cluster. Comparison
with the observed baryonic mass $M_b$ again shows that
$M \gg M_b$.  In particular, observations using the x-ray
satellites XMM-Newton and Chandra indicate that the ratio of
baryonic matter to dark matter in clusters is ${\Om_b}/{\Om_m}
\sim {1}/{6}$.  Additional constraints to the cluster-gas
distribution can be obtained from the Sunyaev-Zeldovich (SZ)
effect.  This is the upscattering of cosmic microwave background
(CMB) photons by hot electron
gas in the cluster; the magnitude of the observed
CMB-temperature change is then proportional to the integral of
the electron pressure through the cluster  (see, e.g.,
\cite{Sunyaev:1980vz}).

\medskip
\noindent {\sl Exercise 4. Estimate, in order of magnitude, the
     x-ray luminosity $L_{{\rm X}}$ for a cluster with total
     mass $M \sim 10^{15} M_{\odot}$ and a baryon fraction 1/6
     in hydrostatic equilibrium with maximum radius $R \sim {\rm Mpc}$.}  
\medskip

\noindent {\sl Exercise 5. Assume the cluster in Exercise 4. is
     isothermal ($T(r) =T =$\, const.) with a dark-matter
     distribution consistent with an NFW profile with $r_c\simeq
     R/10$.   Neglecting the self-gravity of the gas:
\begin{itemize}
\item (a) Show the properly normalized dark-matter density
     profile is approximately $\rho(r) \simeq (233/45)
     M_c/[r(r+r_c)^2]$, where $M_c=M_{<}(r_c)$ is the mass enclosed
     within the scale radius $r_c$.  Determine $M_{<}(r)$ and $M_c$
     and in terms of $M$ for this cluster. 
\item (b) Using your results from (a) solve Eq.~\ref{euler} and
     show that the gas density profile in such an NFW cluster takes
     the form $\rho_{gas}(r) \propto (1+r/r_c)^{\Gamma r_c/r}$, where
     $\Gamma \propto (G M_c \mu m_p/r_c)/(k_B T)$.
\end{itemize}
}
\medskip

\subsubsection{Dynamics}

According to the virial theorem, the velocity dispersion of galaxies is
approximately $v^2(r) \sim G  M_{<}(r)/r$, where $M_{<}(r)$ is
the mass enclosed within a radius $r$.  
Therefore, from measurements  of the velocity dispersion and
size of a cluster (which can be determined if the redshift and
angular size of the cluster are known), one can infer the total
mass $M$.  Once again, the total mass is much larger than the
baryonic mass $M \gg M_b$.

Cluster measurements are by now well established, with many
well-studied and very well-modeled clusters, and there is a good
agreement of estimates of $M$ from dynamics, lensing, X-ray
measurements, and the SZ effect.  The current state of the art
actually goes much further: one can now not only establish the
existence of dark matter, but also map its detailed distribution
within the cluster.

\medskip
\noindent {\sl Exercise 6. Following Zwicky \cite{coma1937}, use
     the virial theorem to find an approximate formula relating the
     average mass of a galaxy in a galaxy cluster to the observed size
     and velocity dispersion of the cluster assuming that the system
     is self-gravitating (and assuming only that the observed
     galaxies contribute to the mass of the system).  What
     answer would Zwicky have found for the Coma cluster with
     modern data?}
\medskip

\subsection{Cosmic Microwave Background and Large-Scale
Structure}

\begin{figure}[t]
  \centering
  \includegraphics[scale=0.6]{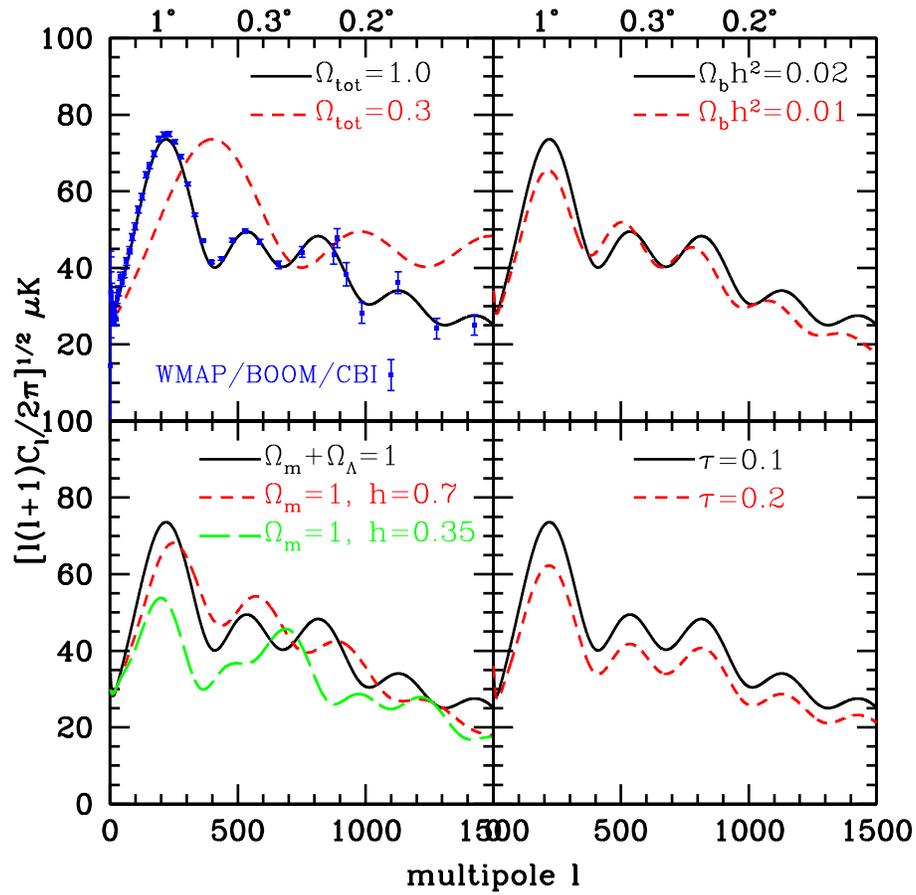}
  \caption{Dependence of the CMB power spectrum on the
  cosmological parameters. From Ref.~\protect\cite{Kam07}.}
  \label{fig:Cls}
\end{figure}

Measurements of the cosmic microwave background (CMB) radiation
and large-scale structure (LSS) of the Universe provide
perhaps the most compelling evidence that the dark matter is
non-baryonic and the most precise measurements of its
abundance.

One obtains from CMB maps the angular power spectrum $C_\ell$ of CMB
temperature anisotropies as a function of multipole
$\ell$.  If the temperature $T(\hatn)$ is measured
as a function of position $\hatn$ on the sky, then one can
obtain the spherical-harmonic coefficients $a_{\ell m} = \int
d\hatn T(\hatn) Y_{\ell m}^*(\hatn)$.  The $C_\ell$'s are then
simply the variance of the spherical-harmonic coefficients:
$C_\ell = \langle |a_{\ell m}|^2\rangle$.  Theoretical predictions for
the power spectrum depend on the values of cosmological
parameters like the matter density $\Om_m 
h^2$, the baryon density $\Om_b h^2$, the cosmological constant
$\Lambda$, the scalar spectral index $n_s$, the optical depth
$\tau$ due to reionization, and the Hubble parameter $H_0$.
One can thus determine these cosmological parameters by fitting
precise measurements of the $C_\ell$s to the theoretical
predictions~\cite{Jungman:1995bz}.  Current measurements provide
detailed information on $C_\ell$ over the range $2 <l<{\cal
O}(1000)$, thus providing precise constraints to the
cosmological parameters.

In the year 2000, data from the Boomerang and MAXIMA experiments
(with supernova measurements) gave $\Om_m
h^2 = 0.13 \pm 0.05$ with error bars that shrink to
$\pm 0.01$ taking into account other measurements or assumptions
(e.g., LSS, Hubble-constant, and supernova measurements, and/or
the assumption of a flat Universe) \cite{Jaffe:2000tx}.  Now,
with WMAP, $\Om_m h^2 = 0.133 \pm 0.006$ and $\Om_b h^2 = 0.0227
\pm 0.0006$ \cite{wmap5}.

\medskip
\noindent {\sl Exercise 7. Suppose that the temperature is
     measured with a Gaussian noise $\sigma_T\simeq
     25~\mu{\mathrm K}$ in $N_{\rm pix} \sim 10^6$ pixels on the
     sky.  Estimate the rms temperature $\left\langle
     \left(\del T/T \right)^2 \right\rangle^{1/2}$ that results.}
\medskip

\section{Basic properties of dark matter}

Having established the existence of dark matter and presented
the case that it is nonbaryonic, we now consider the
requirements for a dark-matter candidate and discuss some
possibilities.  Every dark-matter candidate should satisfy
several requirements:
\begin{itemize}
\item Dark matter must be \emph{dark}, in the sense that it must generically
have no (or extremely weak) interactions with photons;
otherwise it might contribute to the dimming of quasars, create absorption lines in the spectra of distant quasars \cite{Profumo:2006im}, or emit
photons.  One way to  quantify this is by assuming that dark-matter
particles have a tiny charge $fe$ (where $e$ is the electron
charge and $f\ll1$), which can be quantitatively constrained
\cite{Davidson:2000hf}.

\item Self-interactions of the dark matter should be small. We
can estimate the cross section for DM-DM scattering in the
following way: if DM particles scatter less
than once in the history of the Universe, then the mean free
path is less than $\lam = v_{DM} H_0^{-1} \sim \left( 3 \times 10^7
\cm/\sec \right) \left( 10^{17} \, \sec \right) \sim 3
\times 10^{24} \, \cm$. Then, if the galactic-halo density is
$\rho_{DM} \sim 10^{-24}
\textrm{g}/\cm^3$, the opacity for self-scattering in the
galactic halo is
$\kappa = (\rho_{DM} \lam)^{-1} = \sig/m \sim
\cm^2/\textrm{g}$.  Thus, if the elastic-scattering cross
section is $\sig \gtrsim 10^{-24}\,(m/\textrm{GeV}) \, \cm^2 $,
then $\kappa\gtrsim1$ and the typical halo--dark-matter particle
scatters more than once during the history of the Universe.
If dark matter self-scattered, it would suffer
\emph{gravothermal catastrophe}: that is, in binary
interactions of two dark-matter particles, one particle can get
ejected from the halo, while the other moves to a lower-energy
state at smaller radius.  As this occurs repeatedly much of the halo
evaporates and the remaining halo shrinks.  Although a variety of
arguments can constrain dark-matter self-interactions, stringent
and very transparent constraints come from observations of the
Bullet Cluster, the merger of two galaxy clusters, in which it
is seen (from gravitational-lensing maps of the projected matter
density) that the two dark-matter halos have passed through each
other while the baryonic gas has shocked and is located between
the two halos \cite{Randall:2007ph}.

\item Interactions with baryons must also be weak.
Suppose baryons and dark matter interact.  As an overdense
region collapses to form a galaxy, baryons and dark
matter would fall together, with photons radiated from this
baryon-DM fluid.  This would result in a baryon-DM disk, in
contradiction with the more diffuse and extended dark-matter
halos that are observed.  If DM interacted with baryons other
than gravitationally in the early Universe, the 
baryon-photon fluid would be effectively heavier (have a higher
mass loading relative to radiation pressure) even before
recombination, so that the baryon acoustic oscillations in the
matter power spectrum and the CMB angular power spectrum would
be modified \cite{Sigurdson:2004zp}.

\item Dark matter cannot be made up of Standard Model (SM) particles,
since most leptons and baryons are charged.  The only potentially
suitable SM candidate
is the neutrino, but it cannot be dark matter because of the
celebrated Gunn-Tremaine bound~\cite{GT}, which imposes a lower
bound on the masses of dark-matter particles that decoupled when relativistic.  The argument is
the following:  The momentum distribution in the Galactic halo is roughly
Maxwell-Boltzmann with a momentum uncertainty $\Del p \sim m_\nu
\langle v \rangle$ ($\langle v \rangle \sim 300\,
\textrm{km/sec}$), while the mean spacing between neutrinos is
$\Del x \sim n_\nu^{- 1/3} \sim \left(\rho_\nu/m_\nu
\right)^{- 1/3}$.  The Heisenberg uncertainty principle
gives $\Del x \, \Del p \gtrsim \hbar$, which translates into
a lower bound $m_\nu \gtrsim 50\, \textrm{eV}$.  (This
Heisenberg bound can actually be improved by a factor of 2 by
using arguments involving conservation of phase space.) Stronger bounds
($m_\nu \gtrsim 300\, \textrm{eV}$) can be obtained from dwarf
galaxies which have higher phase-space densities.  
As discussed below, there will be a cosmological density of
neutrinos left over from the big bang, with a density $\Om_{\nu}
h^2 \sim 0.1\, (m_\nu/10\, \textrm{eV})$.  The neutrinos of
mass $m_\nu \gtrsim 300\, \textrm{eV}$ consistent with the
Gunn-Tremaine bound would overclose the Universe.  Thus, neutrinos
are unable to account for the dark matter.
\end{itemize}

\section{Weakly Interacting Massive Particles (WIMPs)}

Perhaps the most attractive dark-matter candidates to have been
considered are weakly-interacting massive particles
(WIMPs).  Many theories for new physics at the electroweak scale
(e.g., supersymmetry, universal extra dimensions)
introduce a new stable, weakly-interacting particle, with a mass
of order $M_\chi \sim 100 \, \GeV$.

For example, in supersymmetric (SUSY) theories, the WIMP is the neutralino
\begin{equation}
     \tilde{\chi} = \xi_\gam \tilde{\gam} + \xi_Z \tilde{Z}^0 +
     \xi_{h1} \tilde{h}^0_1+\xi_{h2} \tilde{h}^0_2,
\end{equation}
a linear combination of the supersymmetric partners of the
photon, $Z^0$ boson, and neutral Higgs bosons. Neutralinos are
neutral spin-$1/2$ Majorana fermions.  In theories with
universal extra dimensions there are Kaluza-Klein (KK) states
$\gam_{KK}$, $Z^0_{KK}$, $H^0_{KK}$, which are neutral KK bosons.
The candidates are stable
(or quasi-stable; i.e., lifetimes greater than the age of the
Universe $\tau \gg t_U$) and particle-theory models
suggest masses $M_\chi \sim 10 - 10^3 \, \GeV$.

\begin{figure}[t]
  \centering
  \includegraphics[scale=1.0]{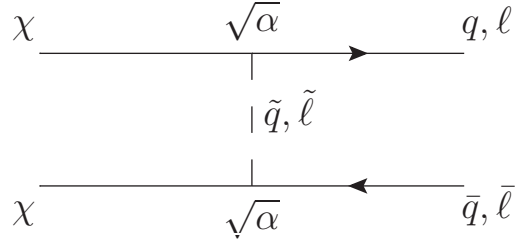}
  \caption{An example of a Feynman diagram for annihilation of
  two WIMPs $\chi$ (neutralinos in this case) to
  fermion-antifermion pairs (where the fermions are either
  quarks $q$ or leptons $l$) via exchange of an
  intermediate-state squark $\tilde q$ or slepton $\tilde l$.}
	\label{fig:annihilation}
\end{figure}

In typical theories two WIMPs can annihilate to SM
particles. For example, for a neutralino we have
the tree-level diagram in Fig.~\ref{fig:annihilation},
where $m_{\tilde{q}, \tilde{l}} \sim 100 \, \GeV$,
so that $\sig \sim \al^2 m_{\tilde{q}, \, \tilde{l}}^{-4}
M_\chi^2 \sim 10^{-8} \, \GeV^{-2}$.

\subsection{WIMP Freezeout in Early Universe}

We now estimate the relic abundance of WIMPs in the standard
scenario of thermal production (see, e.g.,
Ref.~\cite{KolbTurner}).  In the early Universe, at temperatures
$T\gg M_\chi$, WIMPs are in thermal equilibrium and are
nearly abundant as lighter particles, like photons, quarks,
leptons, etc.  Their equilibrium abundance is maintained via
rapid interconversion of $\chi\chi$ pairs and
particle-antiparticle pairs of Standard Model particles.  When
the temperature falls below the WIMP mass, however, the WIMP
abundances become Boltzmann suppressed, and WIMPs can no longer
find each other to annihilate. The remaining WIMPs constitute a
primordial relic population that still exists today.

We now step through a rough calculation.  To do so, we assume
that the WIMP is a Majorana particle, its own antiparticle (as
is the case for the neutralino, for example), although the
calculation is easily generalized for WIMPs with antiparticles
(e.g., KK WIMPs).

The annihilation rate for WIMPs is $\Gam(\chi \chi
\leftrightarrow q \bar{q} , \, \ell \bar{\ell} , \, \dots) =
n_\chi \langle \sig v \rangle$, where $\sigma$ is the cross
section for annihilation of two WIMPs to all lighter
standard-model particles, $v$ is the relative velocity, and the
angle brackets denote a thermal average.  The expansion rate of the
Universe is $H = \left( 8 \pi G \rho/3 \right)^{1/2} \sim
T^2/M_{\mathrm{Pl}}$ during the radiation era, where $\rho \propto T^4$.
In the spirit of  ``squiggly lines'' we have neglected factors
like the effective number of relativistic degrees of freedom
$g_*$ in the expansion rate, which the careful reader can
restore for a more refined estimate.

\begin{figure}[h]
  \centering
  \includegraphics[scale=0.5]{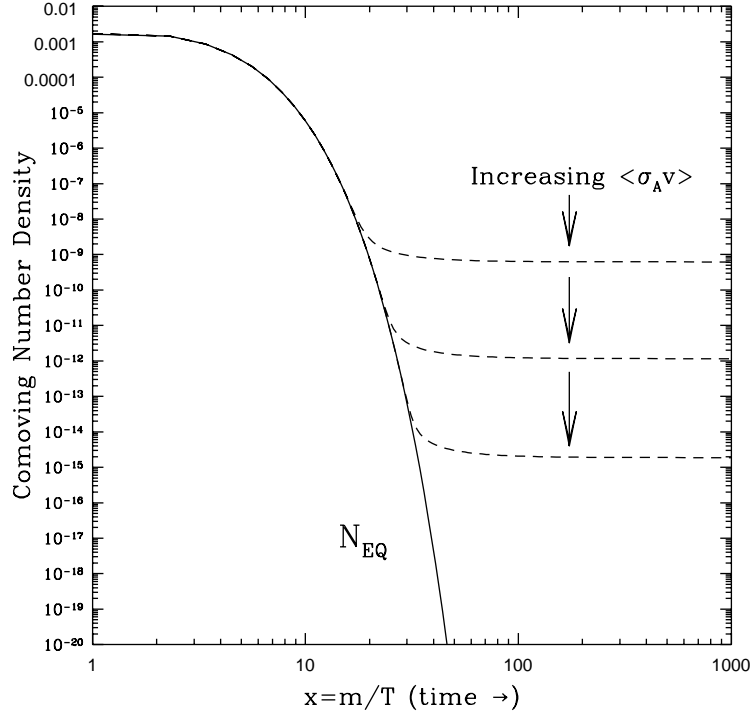}
  \caption{Equilibrium (solid curve) and relic abundance (dashed
  curves) of WIMP particles.
  From Ref.~\protect\cite{Jungman:1995df}.}
	\label{fig:freezeout}
\end{figure}

By comparing these two rates, one can identify two different regimes:
\begin{itemize}
 \item At early times, when $T \gg M_\chi$, $n_\chi \propto T^3$
 and $\Gam \gg H$: particles scatter and annihilate many times
 during an Hubble time and this maintains chemical equilibrium.
\item At late times, when $T \ll M_\chi$, $n_\chi \propto T^{3/2}e^{- M_\chi/T}$
   (note that the chemical potential $\mu_X = 0$ in the case of
   Majorana particles such as the neutralino) and $\Gam \ll H$:
   there can be no annihilations, and the WIMP abundance freezes
   out (the comoving number density becomes constant).
\end{itemize}
This sequence of events is illustrated in Fig.~\ref{fig:freezeout},
which shows the comoving number density of WIMPs as a function
of the inverse temperature in equilibrium (solid curve) and
including freezeout (dashed curves).

Freezeout occurs roughly when $\Gam(T_f) \sim H(T_f)$.
For nonrelativistic particles, $n_\chi = g_\chi \left(
M_\chi T/2 \pi \right)^{3/2}
e^{-M_\chi/T}$, so the freezeout condition becomes
\begin{equation}
      \left( M_\chi T_f \right)^{3/2} e^{-M_\chi/T_f} \sim
      \frac{T_f^2}{M_{Pl}}\quad  \ergo \quad
       \frac{T_f}{M_\chi} \sim \ln \left[ \frac{M_{Pl} M_\chi^{3/2}
        \langle \sig v \rangle}{T_f^{1/2}} \right],
\end{equation}
where the mass parameters are in GeV.
Taking $\langle \sig v \rangle \sim \al^2/M_\chi^2$, and
taking as a first guess $T_f \sim M_\chi$, we finally find
\begin{equation}
      \frac{T_f}{M_\chi} \sim \left\{ \ln \left[ \frac{M_{Pl}
       \al^2}{(M_\chi T_f)^{1/2}} \right] \right\}^{-1} \sim
        \left\{ \ln \left[ \frac{10^{19} 10^{-4}}{100} \right]
         \right\}^{-1} \sim \frac{1}{25} + \textrm{log corrections},
\end{equation}
where the numerical values are characteristic electroweak-scale
parameters (i.e. $\sig \sim 10^{-8} \, \GeV^{-2}$, $M_\chi \sim
100 \,\GeV$).

At freezeout, the abundance relative to photons is
\begin{equation}
  \frac{n_\chi}{n_\gam} = \frac{\Gam(T_f)/{\langle \sig v
  \rangle}}{T_f^3} = \frac{H(T_f)/{\langle \sig v \rangle}}{T_f^3}
  \sim \frac{T_f^2}{M_{Pl} \langle \sig v\rangle T_f^3}
  \sim \frac{1}{M_{Pl} \langle \sig v \rangle T_f} \sim
  \frac{25}{M_{Pl} \langle \sig v \rangle M_\chi}.
\end{equation}
Today we know that
\begin{equation}
  \Om_\chi = \frac{\rho_\chi}{\rho_c} \sim
  \frac{n_\chi^0}{n_\gam^0} \frac{M_\chi n_\gam^0}{\rho_c}
  \sim \frac{25}{M_{Pl}\langle \sig v \rangle} \frac{400 \,
  \cm^{-3}}{10^{-6} \, \GeV \, \cm^{-3}} \, ,
 \end{equation}
with no explicit dependence on the particle mass.

We thus obtain the observed abundance $\Om_\chi h^2 \sim 0.1$ for
$\sig \sim 10^4\, (0.1 \times 10^{19} \times 10^{-6})^{-1} \,
\GeV^{-2} \sim 10^{-8} \, \GeV^{-2}$
which turns out to be nearly exact, even though
we have been a bit sloppy.  A more precise calculation
(including all the factors we have dropped) gives
\begin{equation}
  \Om_\chi h^2 \sim 0.1 \left( \frac{3 \times 10^{-26} \,
  \cm^3/\sec}{\langle \sig v \rangle} \right) + \textrm{ log corrections},
\label{eqn:omega}
\end{equation}
a remarkable result, as it implies that if there is a new
stable particle at the electroweak scale, it is the dark matter.

As an aside, note that partial-wave unitarity of annihilation
cross sections requires $\sig \lesssim M_\chi^{-2}$,
which means $\Om_\chi h^2 \gtrsim \left( M_\chi/300 \, \TeV \right)^2$.
This thus requires $\Om_\chi h^2 \lesssim 0.1$, $M_\chi \lesssim
100 \, \TeV$, without knowing anything about particle physics
\cite{Griest:1989wd}.  More precisely, this bound applies for
point particles and does not apply if dark matter particles are
bound states or solitons.  If the interactions are strong, $\al
\sim 1$, the bound is already saturated.

Although our arguments have been rough, one finds in SUSY and KK
models that there are many combinations of reasonable values for
the the SUSY or KK parameters that provide a WIMP with $\Om_\chi
h^2 \sim 0.1$ for $10 \, \GeV \lesssim M_\chi \lesssim 1 \, \TeV$.

\medskip
\noindent {\sl Exercise 8. Eq.~(\ref{eqn:omega}) was derived
     assuming that the annihilation cross section $\ens{\sig v}$
     is temperature-independent.  Redo the estimate for
     $\Omega_\chi h^2$ assuming that $\ens{\sig v} \propto T^n$,
     where $n=1,2,3,\cdots$.}
\medskip

\subsection{Direct detection}

\begin{figure}[h]
  \centering
  \includegraphics[scale=0.7]{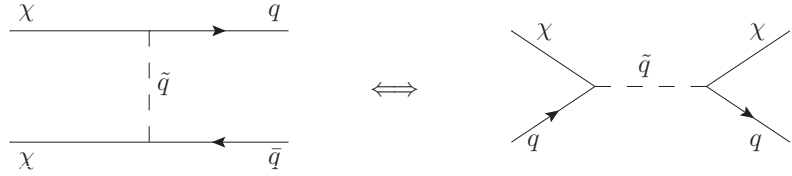}
  \caption{Crossing symmetry between annihilation and scattering diagrams.}
	\label{fig:crossing}
\end{figure}

If WIMPs make up the halo of the Milky Way, then they have a
local spatial density $n_\chi \sim
0.004\,(M_\chi/100\,\GeV)^{-1} \cm^{-3}$ (roughly one per
liter), and are moving with velocities
$v\sim200$~km~sec$^{-1}$.  
Moreover, there is a crossing symmetry between the annihilation
$\chi \chi \to q \bar{q}$ and the elastic scattering $\chi q \to
\chi q$ processes---apart from some kinematic factors the diagrams
are more or less the same (as shown in
Fig.~\ref{fig:crossing})---so one expects roughly that the cross
section $\sig(\chi q \to \chi q) \sim \sig(\chi \chi \to q
\bar{q}) \sim 10^{-36} \, \cm^2$.  One can therefore hope to
detect a WIMP directly by observing its interaction with some
target nucleus in a low-background detector composed, e.g., of
germanium, xenon, silicon, sodium, iodine, or some other element.

At low energies, quarks are bound into nucleons and nucleons
in turn are bound into nuclei, so the cross section one actually
needs is $\sig(\chi N \to \chi N)$ (where $N$ here stands for a
nucleus).  The calculation relating the $\chi q$ interaction to
the $\chi N$ interaction requires both QCD and nuclear physics.
It is complicated but straightforward.  Here we will simply
assume, for illustration, that $\sig(\chi N \to \chi N) \sim
\sig(\chi q \to \chi q)$.

The rate at which a nucleus in the detector is hit by halo WIMPs is then
\begin{equation}
  R \sim n_\chi \sig v \sim (0.004 \, \cm^{-3}) (10^{-36} \,
  \cm^2) \left(3 \times 10^7 \frac{\cm}{\sec} \right) \sim
  10^{-24} \textrm{yr}^{-1};
\end{equation}
if there are $6 \times 10^{23}\, M/(A \, \textrm{g})$ nuclei in a detector,
for an atomic number $A \sim 100$ we expect to see $R \sim
10/\textrm{kg}/ \textrm{ yr}$ events. 

Let us estimate the recoil energy of a nucleus struck by a WIMP.
If a WIMP of $M_\chi \sim 100 \, \GeV$ runs into a nucleus with $A \sim 100$,
the momentum change is $\Del p \sim M_\chi v$, and the nucleus
recoils with an energy of order $E \sim (\Del p)^2/2 m \sim
(100 \, \GeV \, 10^{-3})^2(100 \, \GeV)^{-1} \sim 100 \,
\keV$.

To do things more carefully, one has to account for the fact
that the cross section one actually needs are the interaction
cross sections with nuclei, and via the following steps,
\begin{displaymath}
  \sig(\chi q)  \underset{\textrm{QCD}}{\longrightarrow}
  \sig(\chi n), \sig(\chi p) 
  \underset{\textrm{nuclear physics}}{\longrightarrow} \sig(\chi N),
\end{displaymath}
some theoretical uncertainties are introduced.  One also finds
that $\sigma(\chi N)$ is reduced relative to $\sigma(\chi q)$ by
several orders of magnitude.

Qualitatively, there are two different types of interactions, axial and scalar (or spin-dependent and spin-independent).
The first is described by the Lagrangian,
\begin{equation}
  \L_{\textrm{axial}} \propto \bar{\chi} \gam^\mu \gam_5 \chi \,
  \bar{q} \gam_\mu \gam_5 q,
\end{equation}
which couples $\chi$ to the spin of unpaired nucleons;
this works only for nuclei with spin, and the coupling is
different for unpaired protons or neutrons.  Through this
interaction one expects $\sig \propto {\bar{s}}^2$, where
$\bar{s}$ is the average spin $\sim 1/2$ of the unpaired proton
or neutron in nuclei with odd atomic number.

The second interaction is described by the Lagrangian,
\begin{equation}
  \L_{\textrm{scalar}} \propto \bar{\chi} \chi \bar{q} q,
\end{equation}
which couples $\chi$ to the mass of the nucleus, thus giving a
cross section $\sig \propto M^2 \propto A^2$ (where $M$ and $A$
are the nuclear mass and atomic number),
which implies higher cross sections for larger A.
However, this scaling is only valid up to a limit. In fact, the
momentum exchanged is
$\Del p \sim (100 \, \GeV) (10^{-3}) \sim 0.1 \, \GeV$,
and the nuclear radius is roughly $r \sim A^{1/3} 10^{-13} \, \cm$,
so from the uncertainty principle one has $r\, \Del p \gtrsim 1$ when
\begin{equation}
  \frac{(0.1 \, \GeV) (10^{-13} \, \cm)}{2 \times 10^{-14} \,
  \GeV \, \cm} A^{1/3} \gtrsim 1 \, \qquad \Longrightarrow \qquad\,  A
  \gtrsim 10.
\end{equation}
Detailed calculations show that the cross section for
WIMP-nucleus elastic scattering does not increase much past $A
\gtrsim 100$.

In experiments, people usually draw exclusion curves for the
WIMP-nucleon cross section versus the WIMP mass $M_\chi$.
The exclusion curves are less constraining both for low $M_\chi$
because of the low recoil energy,
and for large $M_\chi$ because (for fixed local energy density $\rho_\chi$) the number density $n_\chi
\propto M_\chi^{-1}$.
\begin{figure}[h]
  \centering
  \includegraphics[scale=0.45]{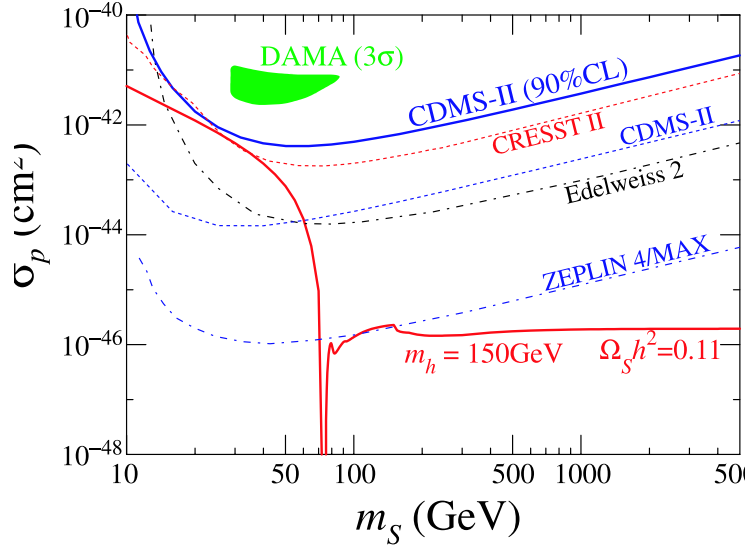}
  \caption{Exclusion plot for the spin-independent dark-matter
  parameter space.  The region favored by the DAMA annual
  modulation is inconsistent with the current bound (solid
  curve) from CDMS.  The broken curves are forecasts for future
  experiments.  We also show, for illustrative purposes only,
  predictions for a WIMP model with a lightest-Higgs-boson
  mass of $m_h=150$ GeV.} 
	\label{fig:exclusionplot}
\end{figure}
To date, only the DAMA experiment has reported a positive signal
\cite{Bernabei:2000qi}.
They used NaI, in which both nuclei have spin, one with an
unpaired proton and the other with an unpaired neutron.  The
interpretation of their signal in terms of a WIMP with scalar
interactions was ruled out by null results (at the time) from
CDMS.  An interpretation of their signal in terms of
a spin-dependent WIMP-neutron interaction was ruled out by the
null search in their Xe detector \cite{Ullio00}.  While the
interpretation in terms of
spin-dependent WIMP-proton scattering was consistent with null
results from other direct searches \cite{Ullio00}, it was ruled
out by null searches for energetic neutrinos from the Sun (see
Fig.~\ref{fig:ullio}), as we explain below.  The interpretation
in terms of spin-dependent scattering is now also ruled out
directly by null results from the COUPP experiment
\cite{Behnke:2008zza}.

\subsection{Energetic $\nu$'s from the Sun}

The escape velocity at the surface of the Sun is $v_{s} \sim 600
\, \textrm{km}/\textrm{s}$, while at the center it is $v_{c} \sim
1300 \, \textrm{km}/\textrm{s}$.  If in passing through the Sun,
a WIMP from the Galactic halo scatters from a nucleus (most
likely a proton) therein to
a velocity less than the escape velocity, then it is
gravitationally trapped within the Sun.  As the
gravitationally-trapped WIMP passes through the Sun
subsequently, it loses energy in additional nuclear scatters and
thus settles to the center of the Sun.  In this way,
the number of WIMPs in the center of the Sun is enhanced.
These WIMPs can then annihilate to standard model particles,
through the same early-Universe processes that set their relic
abundance \cite{Silk:1985ax}.  Decays of the annihilation
products (e.g., $W^+ W^-,
Z^0 Z^0, \tau^+ \tau^-, t \bar{t}, b \bar{b}, c \bar{c}, \dots$)
to neutrinos will produce energetic neutrinos that can escape
from the center of the Sun.  The neutrino energies are $E_\nu
\sim \left[ (1/3)-(1/2) \right] M_\chi \sim 100 \, \GeV$
and so cannot be confused with ordinary solar neutrinos, which
have energies $\sim$MeV.
At night, these neutrinos will move up through the Earth.  If
the neutrino produces a muon through a charged-current
interaction in the rock below a neutrino telescope (e.g.,
super-Kamiokande, AMANDA, or IceCube), the muon may be seen.  In
this way, one can search for these WIMP-annihilation--induced
neutrinos from the Sun.

\begin{figure}[ht]
  \centering
  \includegraphics[scale=0.7]{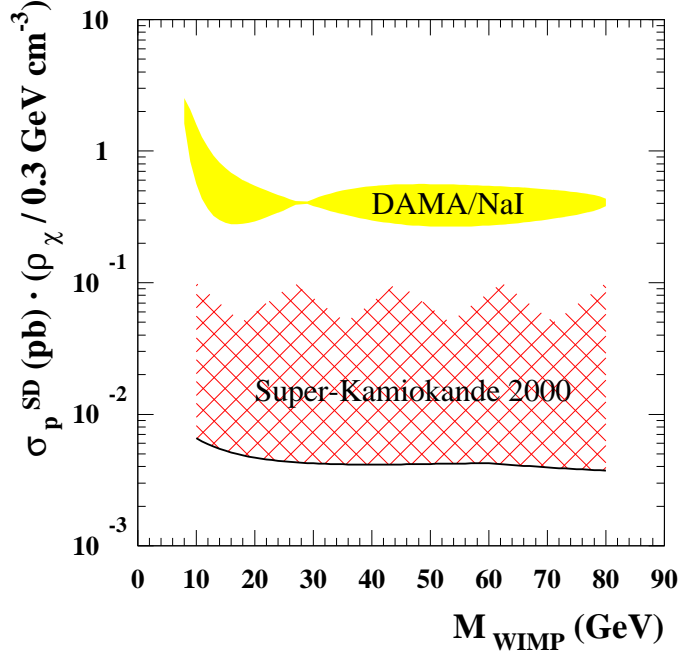}
  \caption{The shaded region shows the parameter space (in
     WIMP mass versus SD WIMP-proton cross section) implied
     by the DAMA annual modulation for a WIMP with exclusively SD
     interactions with protons and no interaction with
     neutrons. The solid curve indicates the upper bound to the
     SD WIMP-proton cross section from null searches for
     neutrino-induced upward muons from the Sun; thus the cross
     hatched region is excluded~\protect\cite{Ullio00}.}
\label{fig:ullio}
\end{figure}

\subsection{Cosmic rays from DM annihilation}

In the Galactic halo, one expects the annihilation processes
$\chi \chi \to \dots \to e^+ e^- , p \bar{p}, \gamma\gamma$;
detection of these products can be a signal of the presence of
dark matter.

\medskip
\noindent {\sl Exercise 9. Show that the annihilation process
     $\chi \chi \to e^+ e^-$ is suppressed for Majorana WIMPs as
     the relative velocity $v \to 0$.} 
\medskip

\subsubsection{Positrons}

\begin{figure}[b]
  \centering
  \includegraphics[scale=0.5]{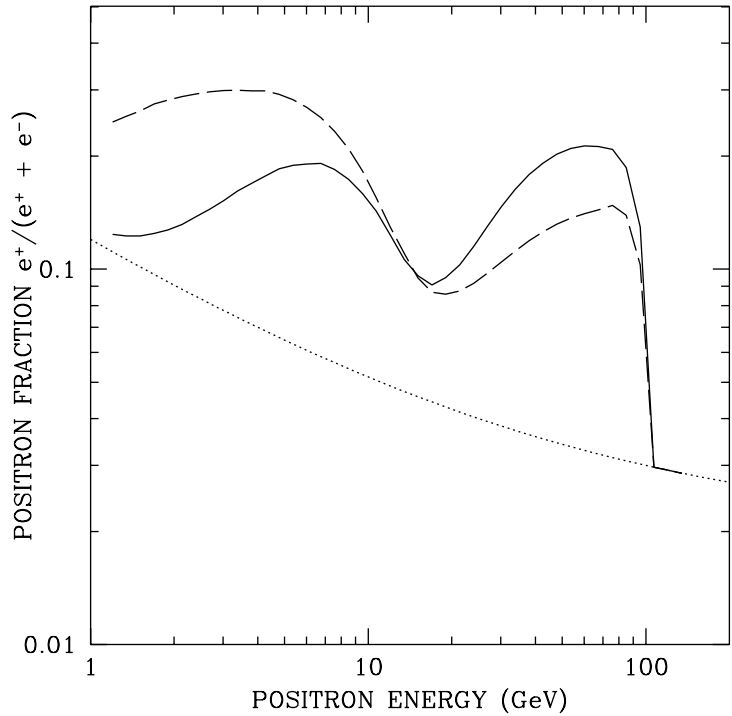}
  \caption{The positron fraction, as a function of
  electron-positron energy, from annihilation of a 120 GeV
  neutralino WIMP to gauge bosons.  The different curves are for
  different cosmic-ray-propagation models, and in both cases,
  the annihilation rate has been boosted by a factor of ten
  relative to the canonical (smooth-halo) value. From
  Ref.~\protect\cite{positrons}.}
	\label{fig:positrons}
\end{figure}

Because of Galactic magnetic fields, cosmic-ray positrons and
antiprotons do not propagate in straight lines and will
thus appear to us as a diffuse background.  Continuum $e^+$'s
from WIMP annihilation are difficult to separate from ordinary
cosmic-ray positrons.  It has been argued that indirect
processes, such as the annihilation $\chi \chi \to W^+ W^- \to
e^+ \nu e^- \bar{\nu}$ \cite{positrons}, will produce a
distinctive bump in the positron spectrum at energies $E_e
\lesssim M_\chi$ (direct annihilation of Majorana WIMPs to
electron-positron pairs is suppressed at Galactic relative
velocities), as illustrated in Fig.~\ref{fig:positrons},
and there has been tremendous excitement recently with the
reported detection by the PAMELA experiment of such a bump
\cite{Adriani:2008zr}.  However, it may be that nearby pulsars
can also produce a bump in the positron spectrum
\cite{Profumo:2008ms}, and more recent results from the Fermi
Telescope \cite{Abdo:2009zk} call the PAMELA result into
questions.  It will thus
be important to understand the possible pulsar signal, as well
as the data, more carefully before the PAMELA excess can be
attributed to WIMP annihilation.

\subsubsection{Antiprotons}

Likewise, it has also been argued that low-energy antiprotons
from WIMP annihilation can be distinguished, through their
energy spectrum, from the more prosaic cosmic-ray antiprotons
produced by cosmic-ray spallation.  Antiprotons can be produced
by the decay of the standard WIMP-annihilation products, and the
energy spectrum of such antiprotons is relatively flat at low
energies.  On the other hand, the energy spectrum of low-energy
cosmic-ray antiprotons due to cosmic-ray spallation decreases at
energies $E\lesssim$GeV.  This is because the process $\bar{p} +
p_{ISM} \to p + p + \bar{p} + \bar{p}$ has an energy threshold,
in the center of mass, of $E_{\mathrm{CM}} > 4 m_p$.  This
requires the primary cosmic-ray momentum to be very high.
Production of an antiproton with $E_{\bar p} \lesssim$GeV
therefore requires that the antiproton be ejected with momentum opposite
to that of the initial cosmic-ray proton, and the phase-space
for this ejection is small.

\subsubsection{Gamma rays}

\begin{figure}[h]
  \centering
  \includegraphics[scale=1.0]{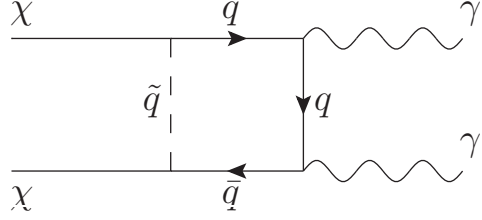}
  \caption{Example of a Feynman diagram for annihilation of two
  neutralinos to two photons through a quark-squark loop.}
  \label{fig:loopgamma}
\end{figure}

A final channel to observe WIMP annihilation is via gamma rays
from WIMP annihilation.  Direct
annihilation of WIMPs to two photons, $\chi \chi \to \gam\gam$,
through loop diagrams such as those shown in Fig.~\ref{fig:loopgamma},
produce monoenergetic photons, with energies equal to the
WIMP mass.  For $v \sim 10^{-3} c$, the photon energies would be
$E_\gam = E_\chi \left( 1 \pm 10^{-3} \right)$, and one would
see a narrow $\gam$-ray line with $\Del \nu/\nu \sim 10^{-3}$,
superposed on a continuum spectrum produced by astrophysical
processes; such a line would be difficult to mimic with
traditional astrophysical sources.  Decays of WIMP-annihilation
products also produce a continuum spectrum of gamma rays at
lower energies.

The other advantage of gamma rays is that they propagate in
straight lines.  This opens up the
possibility to distinguish gamma rays from WIMP annihilation
from those from traditional sources through
directionality---there should be a higher flux of
WIMP-annihilation photons from places where WIMPs are abundant;
e.g., the Galactic center.  Another possibility is dwarf
galaxies, which represent regions of high dark-matter density in
the Milky Way halo.  In general, the $\gam$-ray flux (the number
of photons per unit time-area--solid-angle) is given by
\begin{equation}
  \frac{d F}{d \Om} = \frac{\ens{\sig_{\chi \chi \to \gam \gam}
  v}}{4 \pi M_\chi^2} \int_0^\infty \rho^2(l) dl,
\end{equation}
where the integral is taken along a given line of sight, $l$ is
the distance along that line of sight, and $\rho(l)$ is the
dark-matter density at that distance.
(Note that if $\rho(r) \propto r^{-1}$ with Galactocentric
radius $r$, as in an NFW profile, the intensity formally
diverges, but the flux form any finite angular window around
$r=0$ is finite.)

\medskip
\noindent {\sl Exercise 10. Estimate the $\gam$-ray flux from
     WIMP annihilation, for a given annihilation cross section
     (times relative velocity) $\langle \sigma v
     \rangle_{\mathrm{ann}}$, in an angular window of radius
     $\sim5$ degrees around the Galactic center.  Estimate a
     characteristic $\langle \sigma v \rangle$ for WIMPs and
     evaluate your result for the gamma-ray flux for that value.
      How does it compare, in order of magnitude, with the
      sensitivity of the Fermi Gamma Ray Telescope?}
\medskip

\subsubsection{Galactic Substructure and Boost Factors}

The rate for annihilation, per unit volume, at any point in the
Galactic halo is proportional to $\rho^2$, the square of the
density at that point.  The total annihilation rate in the
halo, or in some finite volume of the halo, is then proportional
to $\int dV\, \rho^2$, the integral, over that volume, of the
density squared.  In the canonical model, the halo density is
presumed to vary smoothly with position in the Galaxy with some
density profile; e.g., the isothermal profile in
Eq.~(\ref{isothermal}).

However, a Galactic halo forms as part of a recent stage in a
sequence of hierarchical structure formation.  In this scenario,
small objects undergo gravitational collapse first; they then
merge to form more massive objects, which then merge to form
even more massive objects, etc.  If some of these substructures
remain partially intact as they merge into more massive halos,
then any given halo (in particular, the Milky Way halo) may have
a clumpy distribution of dark matter.  This is in fact seen in
simulations.  What this implies is that the annihilation rate in
the halo may be enhanced by a ``boost factor'' $B\propto \langle
\rho^2 \rangle/\langle \rho \rangle^2$, where the averages are
over volume in the halo \cite{clumping}.  It may
be possible to see angular variations in the gamma-ray signal
from WIMP annihilation, due to this substructure
\cite{anisotropy,Lee:2008fm}.  It has even been suggested that
proper motions of nearby substructures may be visible
\cite{Koushiappas:2006qq}, although Ref.~\cite{ando} disputed
this claim.

As we will see below, the first gravitationally-collapsed
objects in WIMP models have masses in the range $10^{-6}-100$
Earth masses \cite{Profumo}.  These objects may have densities
several hundred times those of the mean halo density today.  If
so, and if these Earth-mass substructures survive intact through
all subsequent generations of structure formation, then the
boost factor $B$ may be as large as several hundred, implying
much larger cosmic-ray fluxes than the canonical model predicts.

\begin{figure}[h]
  \centering
  \includegraphics[scale=0.5]{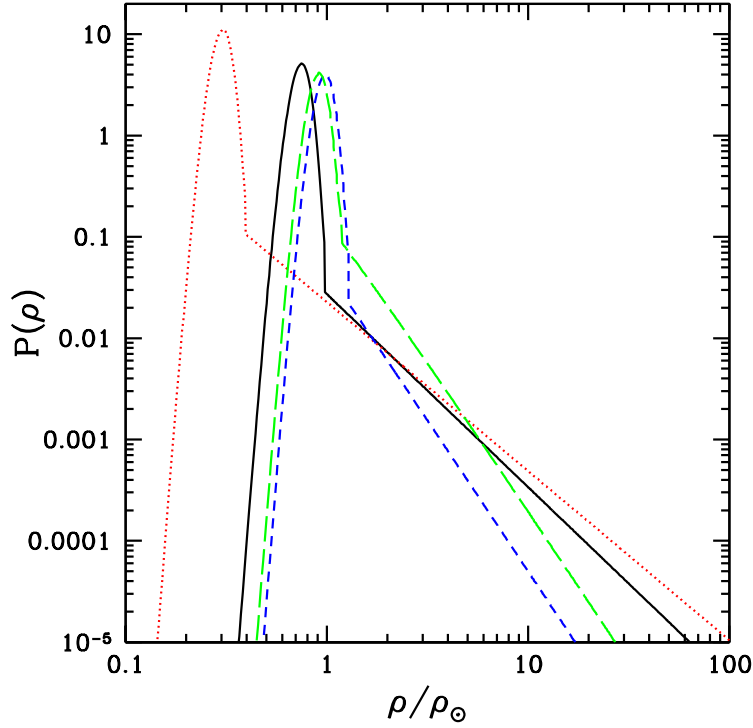}
  \caption{The probability distribution function $P(\rho)$, due
  to substructure, for the local dark-matter density $\rho$, due
  to substructure, in units of the local halo density for a
  smooth halo.  The different curves are for different
  substructure-survival fractions.  The power-law tail is due to
  substructures.  From Ref.~\protect\cite{savvas}.}
  \label{fig:pdf}
\end{figure}

Such large boost factors are, however, unlikely.  Simulations
of recent generations in the structure-formation hierarchy show
that while the tightly bound inner parts of halos may survive
during merging, the outer parts are stripped.
Ref.~\cite{savvas} developed an analytic model, parametrized
in terms of a halo-survival fraction, to describe the
(nearly) scale-invariant process of hierarchical clustering.
This model then provided the boost factor $B$ in terms of that
survival fraction.  By comparing the results (cf.,
Fig.~ \ref{fig:pdf}) of the analytic model for the local
halo-density probability distribution function with subsequent
measurements of the same distribution in simulations (Fig.~1 in 
Ref.~\cite{Vogelsberger:2008qb}), one infers a small
halo-survival fraction.  The analytic model of
Ref.~\cite{savvas} then suggests for this survival fraction no
more than a small boost factor, $B\lesssim$few.

\section{Variations and additions}

What we have described so far may be referred to as the
minimal-WIMP scenario.  In this scenario, the dark matter is a
thermal relic with electroweak-scale cross sections.  It is
neutral and scatters from baryons with cross sections
$\sim10^{-40}\,\cm^2$ (to within a few orders of magnitude).  It
has no astrophysical consequences in the post-freezeout Universe
beyond its gravitational effects.  However, the recent
literature is filled with a large number of astrophysical
anomalies for which explanations have been proposed in terms of
non-minimal WIMPs, WIMPs endowed with extra interactions or
properties.  This is a vast literature, far too large to review
here.  We therefore provide here only a brief sampling,
focusing primarily on those that we have worked on.

\subsection{Enhanced relic abundance}

The calculation above of the freezeout abundance is the standard
one in which it is assumed that the Universe is
radiation-dominated at $T_f \sim 10 - 100 \, \GeV$.  However, we
have no empirical constraints to the expansion rate before big bang
nucleosynthesis, which happens later, at $T_{BBN} \sim 1 \,
\MeV$.

One can imagine other scenarios in which the WIMP abundance changes.
For instance, suppose the pre-BBN Universe is filled with some
exotic matter which has a stiff equation of state, $p_s =
\rho_s$.  This results in a scaling of the energy density of
this stuff $\rho_s \propto a^{-6}$ with scale factor $a$
\cite{relicabundance}.  Such an equation of state may arise if
the energy density is dominated by the kinetic energy of some scalar
field.  The equation of motion of a scalar field with a flat
potential is
\begin{equation}
  \ddot{\varphi} + 3 H \dot{\varphi} = 0 \, \qquad \Longrightarrow \qquad\,
  \ln \dot{\varphi} \propto - 3 \ln a
\end{equation} 
which means that
\begin{equation}
  \rho = \frac{1}{2} \dot{\varphi}^2 \propto a^{-6} \, .
\end{equation}
A stiff equation of state, or something that behaves effectively
like it, may also arise, for example, in scalar-tensor theories
of gravity or if there is anisotropic expansion in the early Universe.

Big-bang nucleosynthesis constrains the energy density of some new
component of matter at a temperature $T\sim$MeV to be
$(\rho_6/\rho_\gam) \lesssim 0.1\, \left(T/\MeV\right)^2$.
Since $\rho_s/\rho_{\mathrm{rad}} \propto T^2$, the expansion
rate with this new stiff matter will at earlier times be $H(T)
\lesssim H_{\mathrm{st}}(T) \left(T/\MeV\right)$, where
$H_{\mathrm{st}}(T)$ is the standard expansion rate.  Neglecting the
logarithmic dependence of the freezeout temperature $T_f \propto
\ln[H \rho_6 n_\gam]$ on the expansion, the WIMP abundance with
this new exotic matter will be
\begin{equation}
  \frac{n_\chi}{n_\gam} = \frac{1}{n_\gam} \frac{\Gam}{\sig v} =
  \frac{1}{n_\gam} \frac{H}{\sig v}
  \lesssim \left( \frac{n_\chi}{n_\gam} \right)_{st} \left(
  \frac{T}{\MeV}\right)
  \sim \left( \frac{n_\chi}{n_\gam} \right)_{st} \left(
  \frac{M_\chi/25}{\MeV}\right).
\end{equation}
Thus, for example, the relic abundance of an $M_\chi \sim 150 \, \GeV$
WIMP can be increased by as much as $\sim10^4$ in this way
\cite{relicabundance,Profumo:2003hq}.

\medskip
\noindent {\sl Exercise 11. Show that anisotropic expansion gives
     rise to a Friedmann equation that looks like that for a
     Universe with a new component of matter with $\rho \propto
     a^{-6}$. To do so, consider a Universe with metric
     $ds^2=dt^2-[a_x(t)]^2 dx^2 -[a_y(t)]^2 dy^2-[a_z(t)]^2dz^2$,
     with $a_x(t)$, $a_y(t)$, and $a_z(t)$ different, and then
     derive the Friedmann equation for a Universe filled with
     homogeneous matter of density $\rho$.} 
\medskip

\subsection{Kinetic decoupling}

There are two different kinds of equilibrium for WIMPs in the primordial bath.
One is chemical equilibrium, which is maintained by the reactions
\begin{equation*}
  \chi \chi \leftrightarrow f \bar{f} \, ;
\end{equation*}
the other is kinetic equilibrium, maintained by the scattering
\begin{equation*}
  \chi f \leftrightarrow \chi f \, .
\end{equation*}

The first reaction freezes out before the second, since $n_f \gg
n_\chi$, where $f$ is any kind of light degree of freedom.
However, $\sig (\nu \chi \leftrightarrow \nu \chi) \propto
E_\nu^2$ since the $\nu$'s are Yukawa coupled,
and $\sig (\gam \chi \leftrightarrow \gam \chi) \propto E_\gam^2$
since the photons are coupled by
$\eps_{\mu \nu \rho \sig} k^\mu k^\nu \eps^\rho \eps^\sig$ \cite{Chen:2001jz}.
This means that $\Gam(\chi f \leftrightarrow \chi f)$ drops
rapidly and so kinetic freezeout happens
not too much later than chemical freezeout.

\begin{figure}[h]
  \centering
  \includegraphics[scale=0.7]{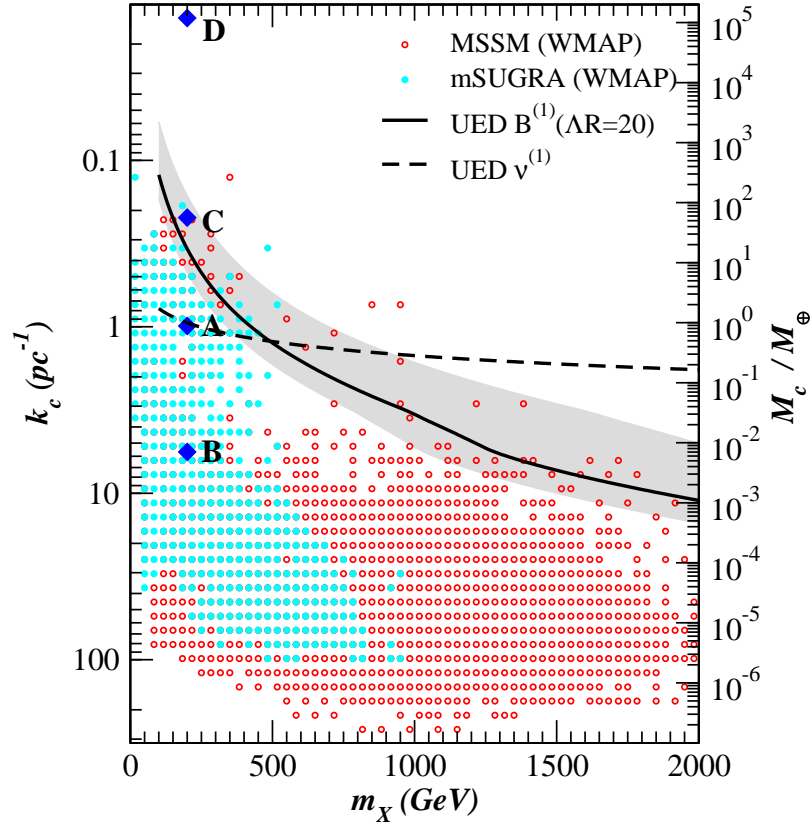}
  \caption{The wavenumber and mass scale at which the primordial
  power spectrum is cut off due to kinetic decoupling of WIMPs
  in supersymmetric and UED models for WIMPs.  From
  Ref.~\protect\cite{Profumo}.}
  \label{fig:kcut}
\end{figure}

Detailed calculations of the kinetic-decoupling temperature
$T_{kd}$ show that $T_{kd}$ varies over 6 orders of magnitude in
scans of the SUSY and UED parameter spaces \cite{Profumo}.
During the time particles are chemically but not kinetically
decoupled, they have the same temperature of the thermal bath,
which scales as $T_\gam \propto a^{-1}$, and after that, $T_\chi
= p_\chi^2/2 M_\chi \propto a^{-2}$.
So, density perturbations $\del \rho_\chi/\rho_\chi$ are
suppressed on $\lam_{phys} \sim H^{-1}$ while the WIMPs are
kinetically coupled.
The cutoff in the power spectrum $P(k)$ is at physical
wavenumber $k_c = H(T_{kd})$, so if $T_{kd}$ decreases,
also $k_c$ decreases.
We expect power suppressed at mass scales $M < M_c$, where $M_c
\sim 10^{-4} - 10^2 M_\oplus$
is the mass enclosed in the horizon at $T_{kd}$, as shown in
Fig.~\ref{fig:kcut} \cite{Profumo}.

\medskip
\noindent {\sl Exercise 12.  Derive the mass $M_{kd}$ enclosed
     within the horizon at a temperature $T_{kd}$.}
\medskip

\subsection{ Particle Decay and Suppression of Small Scale Power}

It might be the case that dark matter is produced by the decay
of a metastable particle that was once in kinetic equilibrium
with the thermal bath.  For instance, although the dark matter
cannot be a charged particle it might be produced by the 
decay of a charged particle.  The growth of perturbation modes
that enter the horizon prior to the decay of the charged
particle will be suppressed relative to the standard case due to
the coupling to the thermal bath: growth of charged-particle
density perturbations is suppressed since charged particles
cannot move through the baryon-photon fluid.  If one has $\chi^+ \to \chi^0
+ e^+$, with $\tau \sim 3.5 \textrm{yr}$ ($z \sim 10^7$), then
the matter power spectrum $P(k)$ is suppressed on $k \gtrsim
\textrm{Mpc}^{-1}$ \cite{Sigurdson:2003vy}, while for shorter
lifetimes structure will be
suppressed for larger $k$ (smaller length scales).  Models
exhibiting charged-particle decay can be found in the parameter
space of standard or minimal extensions of canonical WIMP
(e.g., supersymmetric) scenarios \cite{Profumo:2004qt}.    While
limits on energy injection and the formation of exotic bound
states in big bang nucleosynthesis (BBN) constrain the fraction of
the Universe bound up in charged particles \cite{bbncpd}
the suppression of power due to particle decay in the Universe
remains a potentially observable effect of metastable particles.
It is possible the metastable particle might remain in kinetic
equilibrium via another interaction, or even if the particle is
out of kinetic equilibrium the energy released in the decay
process may impart the dark-matter particle with a velocity high
enough to erase small-scale structure via free streaming
\cite{sssdecay}.

Future measurements of high-redshift cosmic 21-cm fluctuations
may provide a direct probe of modifications to the small-scale
dark-matter power spectrum and other aspects of fundamental
physics (see, e.g., \cite{Profumo:2004qt,cosmic21cm}).

\medskip
\noindent {\sl Exercise 13.  Derive the comoving wavenumber $k$
      that enters the horizon at the time a particle of lifetime
      $\tau$ decays.}
\medskip
 
\subsection{Dipole dark matter}

\begin{figure}[h]
  \centering
  \includegraphics[scale=0.5]{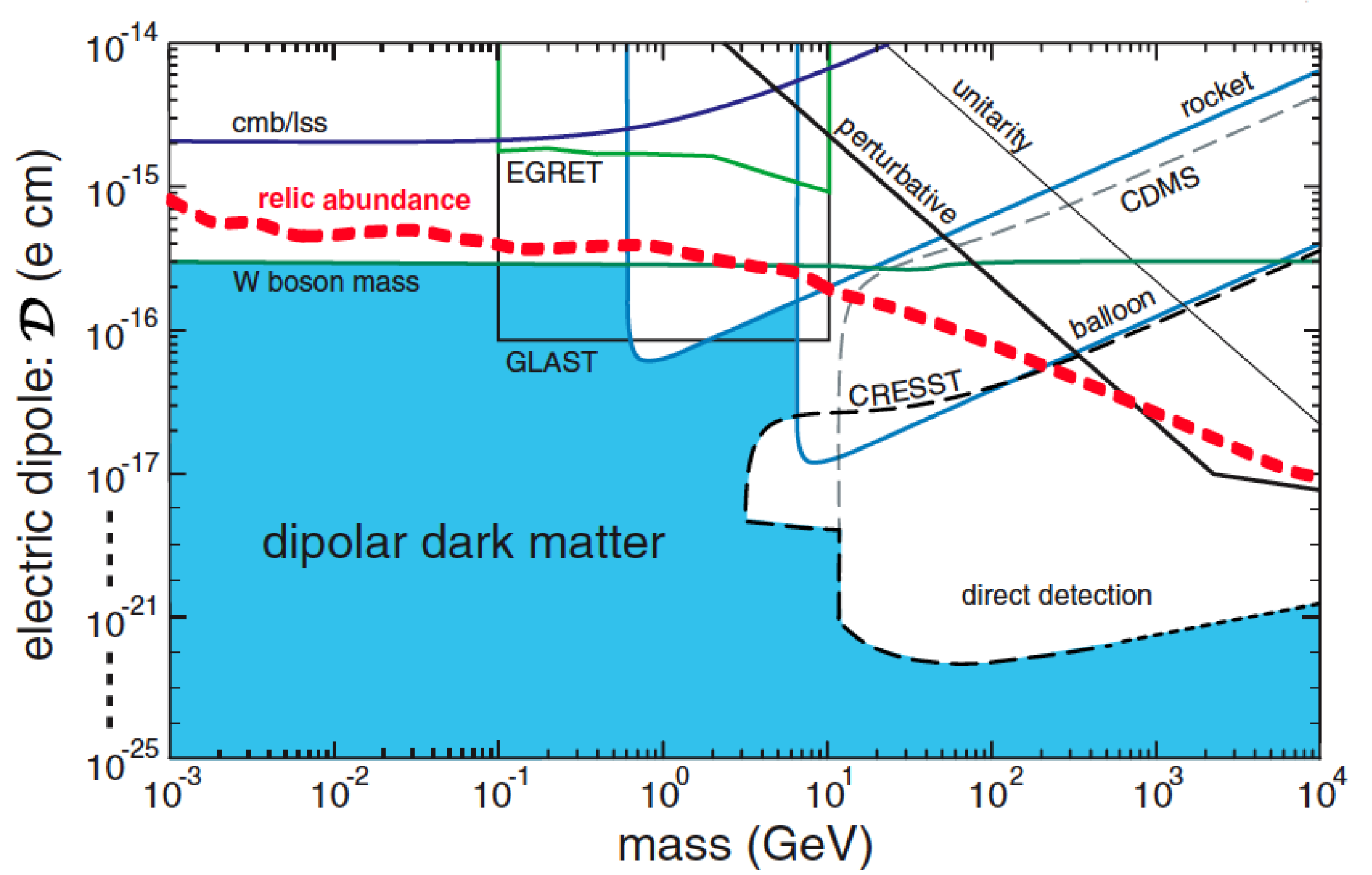}
  \caption{Constraints to the dipole-mass parameter space for
  dark matter with an electric or magnetic dipole.  From
  Ref.~\protect\cite{Sigurdson:2004zp}.}
  \label{fig:dipole}
\end{figure}

While dark matter cannot be a charged particle it may (via
higher order interactions) be endowed with an electric or
magnetic dipole moment interactions of the form
\cite{Sigurdson:2004zp,Profumo:2006im},
\begin{equation}
  \L_{\textrm{dipole}} \propto \bar{\chi_i}
  \sigma_{\mu\nu}\left(\mu_{ij} + \gamma_5 {\cal
  D}_{ij}\right)\chi_j F^{\mu\nu},
\end{equation}
Here, diagonal interaction terms ($i=j$) are the magnetic
($\mu$) or electric (${\cal D}$) dipole moments of a particle
$\chi$, while off-diagonal terms ($i\neq j$) are referred to as
transition moments between the lightest WIMP state $i$ and
another, slightly heavier, WIMP state $j$.  Such a dipole
coupling to photons alters the evolution of dark-matter density
perturbations and CMB anisotropies \cite{Sigurdson:2004zp},
although the strongest
constraints to dipole moments comes from precision tests of the
Standard Model for WIMP masses $M_\chi \lesssim 10$ GeV and
direct-detection experiments for $M_\chi \gtrsim 10$ GeV
\cite{Sigurdson:2004zp,Masso:2009mu}; see Fig.~\ref{fig:dipole}
for the full constraints.

It may be possible to explain the results of the DAMA
experiment using low-mass dipolar dark matter with a transition
moment \cite{Masso:2009mu}.  It may also be
possible to look for the effects of a transition dipole moment
in the absorption of high energy photons from distant sources
\cite{Profumo:2006im}.

\medskip
\noindent {\sl Exercise 14.  Calculate the cross section for
     elastic scattering of a particle with an electric dipole
     moment of magnitude $d$ from a nucleus with charge $Ze$.}
\medskip

\subsection{Gravitational constraints}

It is generally assumed that while dark matter may involve new
physics, the gravitational interactions of the dark matter are
standard.  In other words, it is generally assumed that the
gravitational force between two DM particles and between a
dark-matter particle and a baryon is the same as that between
two baryons.  More precisely, the Newtonian gravitational force
law between baryons that has been tested in the laboratory and
in the Solar System reads $F_{b_1 b_2} = G m_1 m_2/d^2$.
We then usually assume that the force between baryons and DM is
$F_{bd} = G m_b m_d/d^2$, and also that the gravitational DM-DM
force law is $F_{d_1 d_1} = G_d m_{d_1} m_{d_2}/d^2$ with
$G_d=G$.  However, there is no empirical evidence that this is
true at more than the order-unity level \cite{Gradwohl92}, and
it has even been postulated that $G_d =2 G$ in order to account
for the void abundance \cite{Gubser:2004uh}.  A similar behavior (an
increase in the DM-DM force law) could also arise if there were
a new long-range interaction mediated by a nearly massless
scalar field $\vphi$ with Yukawa interactions $\vphi \bar{\psi}
\psi$ with the DM field $\psi$.  The difficulty in providing
empirical constraints to this model is that measurements (e.g.,
gravitational lensing or stellar/galactic dynamics) of the
dark-matter distribution determine only the gravitational
potential $\Phi$ due to the dark-matter distribution,
represented by some density $\rho_d(\vec r)$, obtained through
the Poisson equation $\nabla^2 \Phi = 4 \pi G \rho_d$.  However,
the same $\Phi$ can be obtained by replacing $\rho_d \to
(1/2)\rho_d$ if we simultaneously replace $G \to 2 G$.

It turns out, though, that this exotic interaction can be
constrained by looking at substructures in the Milky Way
halo~\cite{Kesden1,Kesden2}.  The Sagittarius dwarf galaxy,
is dark-matter dominated, and it follows an elongated orbit
around the Milky Way.  When the dwarf reaches its point of
closest approach to the Milky Way, the tidal forces it
experiences in the Milky Way potential are largest.  Stars are
then stripped from the innermost and furthermost edge of the
dwarf.  Those from the innermost parts move at slightly larger
velocities in the Galactic halo and at slightly smaller
Galactocentric radii; they thus subsequently run ahead of the
Sagittarius dwarf and form the leading tidal tail of the
Sagittarius dwarf that is observed.  Conversely, those stripped
from the outer edge subsequently lag behind forming the trailing
tidal tail that is observed.  Observationally, the leading and
trailing tails have roughly the same brightness, as expected.
Suppose now that the DM-DM force law were modified to $G_d = f
G$ with $f>1$.  The dark-matter halo of the Sagittarius dwarf
would then be accelerated toward the Milky Way center more
strongly than the stellar component of the Sagittarius dwarf.
The stellar component would then slosh to the furthermost edge.
Then, when the dwarf reaches its point of closest approach to
the Milky Way, stars are still stripped from the outer edge, forming a
trailing tail.  However, there are now no stars in the innermost
edge to form the leading tail.  The evacuation of stars from the
leading tail is inconsistent with observations, and this leads,
with detailed calculations, to a bound $G_{d} = G \left( 1 \pm
0.1 \right)$ to Newton's constant for DM-DM interactions.  In
other words, dark matter and ordinary matter fall the same way,
to within 10\%, in a gravitational potential well.

While Ref.~\cite{Peebles:2009th} has more recently claimed to run
a simulation of the tidal tails of the Sagittarius dwarf consistent
with $G_d=2G$, Ref.~\cite{Kesden:2009bb} has argued that the initial
conditions for that simulation are self-inconsistent.
Refs.~\cite{Carroll:2008ub,Carroll:2009dw} argue that a new long-range DM-DM
force law implies, under fairly general conditions, a weaker
long-range DM-baryon force law, and they discuss and compare
possible tests of such a scenario.

\subsection{Electromagnetic-like interactions for dark matter?}

Another possibility is that dark matter experiences long-range
electromagnetic-like forces mediated by a dark massless photon
that couples only to gravity.  Of course, if the fine-structure
constant $\alpha_d$ associated with this dark $U(1)$ symmetry is
too large, then long-range dark forces will induce the dark
matter to be effectively collisional.  This constrains
$\alpha_d \lesssim 0.005\, (M_\chi/{\mathrm{TeV}})^{3/2}$
\cite{Ackerman:2008gi}.  Far more restrictive constraints may
arise from the development of plasma instabilities that may
arise if there are (dark) positively and negatively charged
dark-matter particles, but precise calculations of these effects
remain to be done.  See Refs.~\cite{Ackerman:2008gi,Feng:2009mn}
for more discussion of these models.

\medskip
\noindent {\sl Exercise 15.  Estimate the relic abundance of a
     dark-matter particle with dark charge $\alpha_d$ assuming
     that it annihilates to dark-photon pairs and assuming that
     the dark sector has the same temperature as the rest of the
     primordial plasma.}
\medskip

\section{Some other particle dark-matter candidates}

WIMP models are interesting for a number of reasons:  (1) The
correct relic density arises naturally if there is new physics
at the electroweak scale; (2) there are good prospects for
detection of these particles, if they are indeed the dark
matter; and (3) there is synergy with the goals of accelerator
searches (especially at the LHC) for new electroweak-scale
physics.

Still, there are a large number of other particle candidates for
dark matter.  Here we discuss two, the sterile neutrino and the
axion, which may also arise in extensions of the
standard model and for which there are clear paths toward
detection if they make up the dark matter.

\subsection{Sterile Neutrinos}

A convenient mechanism to introduce neutrino masses and explain
their smallness by a minimal extension of the Standard Model is
to add 3 right-handed neutrinos which are singlets under the SM
gauge group.  The mass matrix is taken to be of the form (for
simplicity we consider only one family),
\begin{equation}
\begin{split}
  &\begin{matrix} &\nu_L & \nu_R \end{matrix} \\
  \begin{matrix} \nu_L \\ \nu_R \end{matrix}
  \Bigg( & \begin{matrix}
      &0 & M_D \\
      &M_D & M
      \end{matrix} \Bigg),
\end{split}
\end{equation}
where the $\nu_L$ and $\nu_R$ are left-handed and right-handed
(weakly-interacting and sterile, respectively) fields.

In the see-saw mechanism, the Dirac mass is assumed to be tiny
compared with the Majorana mass: i.e., $M_D \ll M$.  The mass
eigenstates then have masses $M_1 \simeq M_D^2/M \ll M$, and
$M_2\simeq M$.  For our purposes, it is advantageous to map the
two-dimensional $M_D$-$M$ parameter space onto the
$M_s$-$\theta$ parameter space, where $M_s$ is the mass of the
sterile (heavier) neutrino and $\theta$ is the mixing angle
between the two states.  The active and sterile mass eigenstates
can then be written
\begin{align}
  & \ket{\nu_a} = \cos \tht \ket{\nu_L} + \sin \tht \ket{\nu_R}, \\
  & \ket{\nu_s} = - \sin \tht \ket{\nu_L} + \cos \tht \ket{\nu_R},
\end{align}
where $\tht =M_D/M$.

Sterile neutrinos can be produced in the early Universe and have
both (1) a lifetime longer than the age of the Universe and (2)
a cosmological density $\Omega_s\sim 0.2$ if the
sterile-neutrino mass is in the $\sim$keV regime \cite{Dodelson:1993je}.

\begin{figure}[ht]
  \centering
  \includegraphics[scale=1.2]{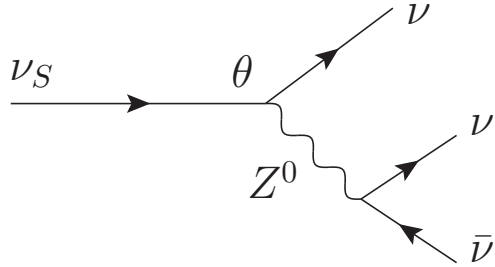}
  \caption{Main decay channel for sterile neutrinos.}
	\label{fig:steriledecay}
\end{figure}

The main decay mode of the sterile neutrino is then $\nu_S \to
\nu \nu \bar{\nu}$, through the exchange of a $Z^0$ boson, as
shown in Fig.~\ref{fig:steriledecay}.  The decay rate and
lifetime are
\begin{equation}
  \Gam = \frac{G_F^2 M_S^5}{96 \pi^3} \tht^2 \qquad \ergo \qquad \tau_S =
  \frac{\hbar}{\Gam} \sim 10^{20} \, \sec \left(
  \frac{M_S}{\keV} \right)^5 \tht^{-2}.
\end{equation}

\begin{figure}[ht]
  \centering
  \includegraphics[scale=1.2]{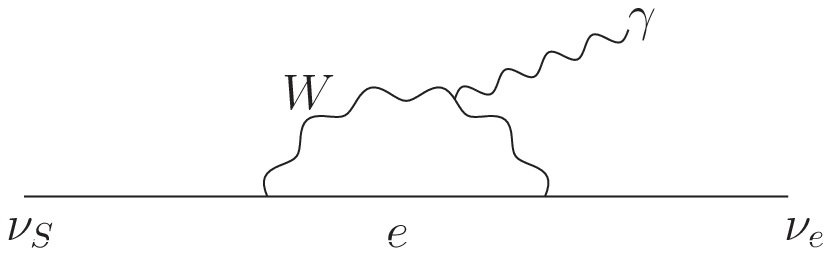}
  \caption{Loop diagram for the decay $\nu_s \to \nu \gam$.}
	\label{fig:sterileloop}
\end{figure}

If the sterile neutrinos constitute the dark matter, then it
must be that $\tau_S \gg 10^{17} \, \sec$, which is possible if
$M_S \sim O(1) \, \keV$.    This mass cannot however be too small,
because of the Gunn-Tremaine limit from dwarf-spheroidal galaxies,
which is $M_S \gtrsim 0.3 \, \keV$.   A stronger constraint to 
the model comes from the X-ray emission in the radiative
decay $\nu_S \to \nu \gam$, through the diagram in
Fig.~\ref{fig:sterileloop}.  This produces an x-ray line that
can be sought in the spectrum of, e.g., a galaxy cluster.  While
null searches for such lines (and from the diffuse cosmic x-ray
background) provide \cite{Xray1,Xray2} stringent constraints to
the model, there are still some regions in the $M_s$-$\theta$
parameter space that remain consistent with current
constraints.  This region may be probed, however, with future
more sensitive x-ray searches. 
One interesting extended application of sterile neutrino dark matter was its use as a potential mechanism for generating momentum-anisotropy during supernova to drive pulsar kicks \cite{Kusenko04}.
See, for instance, Refs.~\cite{Shaposhnikov07,Kusenko:2009up},
for the current status of sterile neutrino dark matter.

\subsection{Axions}

Axions arise in the Peccei-Quinn (PQ) solution to the strong-$CP$
problem \cite{PQ}.  A global $U(1)_{PQ}$ symmetry is
spontaneously broken at a scale $f_a$, and the CP-violating
phase $\theta$ in the QCD Lagrangian becomes a dynamical field with a
flat potential.  At temperatures
below the QCD phase transition, nonperturbative quantum effects
break explicitly the symmetry and produce a non-flat potential
that is minimized at $\theta\rightarrow 0$.
The axion is the pseudo-Nambu-Goldstone boson of this
near-global symmetry, the particle associated with excitations
about the minimum at $\theta=0$.  The axion mass is $m_a \simeq\, {\rm
eV}\,(10^7\, {\rm GeV}/ f_a)$, and its coupling to ordinary
matter is $\propto f_a^{-1}$.

The Peccei-Quinn solution works equally well for
any value of $f_a$.  However, a variety
of astrophysical observations and laboratory experiments
constrain the axion mass to be $m_a\sim10^{-4}$ eV.
Smaller masses would lead to an
unacceptably large cosmological abundance.  Larger masses
are ruled out by a combination of constraints from supernova
1987A, globular clusters, laboratory experiments, and a search
for two-photon decays of relic axions.

Curiously enough, if the axion mass is in the relatively small viable
range, the relic density is $\Omega_a\sim1$, and so the axion may
account for the halo dark matter.  Such axions would be produced
with zero momentum by a misalignment mechanism in the early
Universe and therefore act as cold dark matter.  During the process of
galaxy formation, these axions would fall into the Galactic
potential well and would therefore be present in our halo with a
velocity dispersion near 270 km~sec$^{-1}$.

It has been noted that quantum gravity is generically expected
to violate global symmetries, and unless these Planck-scale
effects can be suppressed by a huge factor, the Peccei-Quinn
mechanism may be invalidated \cite{gravity}.  Of course, we have
at this point no predictive
theory of quantum gravity, and several mechanisms for forbidding
these global-symmetry violating terms have been proposed
\cite{solutions}.  Therefore, discovery of an
axion might provide much needed clues to the nature of
Planck-scale physics.

There is a very weak coupling of an axion to photons through the
triangle anomaly, a coupling mediated by the exchange of virtual
quarks and leptons.  The axion can therefore decay to two
photons, but the lifetime is $\tau_{a\rightarrow \gamma\gamma}
\sim 10^{50}\, {\rm s}\, (m_a / 10^{-5}\, {\rm eV})^{-5}$ which
is huge compared to the lifetime of the Universe and therefore
unobservable.  However, the $a\gamma\gamma$ term in the
Lagrangian is ${\cal L}_{a\gamma\gamma} \propto a {\vec E} \cdot
{\vec B}$ where ${\vec E}$ and ${\vec B}$ are the electric and
magnetic field strengths.  Therefore, if one immerses a resonant
cavity in a strong magnetic field, Galactic axions that pass
through the detector may be converted to fundamental excitations
of the cavity, and these may be observable \cite{sikivie}.  Such
an experiment is currently underway \cite{axionexperiments} and
has already begun to probe part of the cosmologically
interesting parameter space (see the Figure in Ref.~\cite{karlles}), and it
should cover most of the interesting region parameter space in
the next few years. 

Axions, or other light pseudoscalar particles, may show up
astrophysically or experimentally in other ways.  For example,
the PVLAS Collaboration \cite{pvlas} reported the observation of
an anomalously large rotation of the linear polarization of a
laser when passed through a strong magnetic field.  Such a
rotation is expected in quantum electrodynamics, but the
magnitude they reported was in excess of this expectation.  One
possible explanation is a coupling of the pseudoscalar $F \tilde
F$ of electromagnetism to a low-mass axion-like pseudoscalar
field.  The region of the mass-coupling parameter space implied
by this experiment violates limits for axions from astrophysical
constraints, but there may be nonminimal models that can
accommodate those constraints.  Ref. \cite{kris} reviews the
theoretical interpretation and shows how the interactions of
axions and other axion-like particles may be tested with x-ray
re-appearance experiments.  While the original PVLAS results
have now been called into question Ref.~\cite{Chou:2007zzc},
variations of the model may still be worth investigating.

\section{Conclusions}

Here we have reviewed briefly the basic astrophysical evidence
for dark matter, some simple astrophysical constraints to its
physical properties, and the canonical WIMP model for dark
matter.  We then discussed a number of variations of the
canonical model, as well as some alternative particle
dark-matter candidates.  Still, we have only scratched the
surface here, surveying only a small fraction of the
possibilities for non-minimal dark matter.  Readers who are
interested in learning more are encouraged to browse the recent
literature, where they will find an almost endless flow of
interesting possibilities for dark matter, beyond those we have
reviewed here.

\bigskip

\begin{acknowledgement}
We thank Sabino Matarrese for initiating this collaboration
during the Como summer school at which these lectures were
given.  We also thank the Aspen Center for Physics, where part
of this review was completed. This work was supported at Caltech by DoE
DE-FG03-92-ER40701 and the Gordon and Betty Moore Foundation,
and at the University of British Columbia by a NSERC of Canada
Discovery Grant.
\end{acknowledgement}

\end{document}